\documentclass[aps,prl,twocolumn,groupedaddress, superscriptaddress,longbibliography]{revtex4-1}
\usepackage{graphicx}
\usepackage{amsmath}
\usepackage{url}
\usepackage[normalem]{ulem}

\usepackage[colorlinks,bookmarks=false,citecolor=blue,linkcolor=red,urlcolor=blue]{hyperref}

\begin{document}

\title{Resonant light enhances phase coherence \\ in a cavity QED simulator of fermionic superfluidity}
\author{Shane P. Kelly}
\email[Corresponding author:~]{shakelly@uni-mainz.de}
\affiliation{Institut f\"ur Physik, Johannes Gutenberg Universit\"at Mainz, D-55099 Mainz, Germany}
\affiliation{Kavli Institute for Theoretical Physics, University of California, Santa Barbara, CA 93106-4030, USA}
\author{James K. Thompson}
\affiliation{JILA, NIST, Department of Physics, University of Colorado, Boulder, CO 80309, USA}
\author{Ana Maria Rey}
\affiliation{JILA, NIST, Department of Physics, University of Colorado, Boulder, CO 80309, USA}
\affiliation{Center for Theory of Quantum Matter, University of Colorado, Boulder, CO 80309, USA}
\affiliation{Kavli Institute for Theoretical Physics, University of California, Santa Barbara, CA 93106-4030, USA}
\author{Jamir Marino}
\affiliation{Institut f\"ur Physik, Johannes Gutenberg Universit\"at Mainz, D-55099 Mainz, Germany}
\affiliation{Kavli Institute for Theoretical Physics, University of California, Santa Barbara, CA 93106-4030, USA}
\date{\today}
\begin{abstract}
   Cavity  QED  experiments are natural hosts  for non-equilibrium  phases of matter supported by photon-mediated interactions.  
    In this work,  we consider a cavity QED simulation of the  BCS model of superfluidity, by studying regimes where the cavity photons act as dynamical degrees of freedom instead of mere mediators of the interaction via virtual processes. 
    We find  an enhancement of long time coherence following  a quench whenever the cavity frequency is tuned into resonance with the atoms.
    We discuss how this is equivalent to enhancement of non-equilibrium superfluidity and highlight similarities to an analogous phenomena recently studied in solid state quantum optics. 
    We also  discuss the conditions for observing this enhanced resonant pairing in  experiments by including the effect of photon losses and inhomogeneous coupling in our analysis.
\end{abstract}

\maketitle

Superconductivity and superfluidity are among the most celebrated predictions of modern condensed matter theory, both for their fundamental importance and for the promise they hold to revolutionize power transmission~\cite{ashcroft1976solid,annett2004}. 
Recent theory and experimental efforts point at  potential non-equilibrium enhancement of superconducting-like phenomena in platforms at the interface of condensed matter and quantum optics, hinting at novel avenues  beyond conventional high-temperature superconductors   in solid state systems~\cite{PhysRevLett.63.1996, ginsberg1998physical,RevModPhys.66.763}. 
These encompass pump and probe experiments in the solid state setting~\cite{matsunaga2013higgs, matsunaga2014light, mankowsky2014nonlinear,mitrano2016possible,isoyama2021light}, as well as  proposals to enhance superconducting order using driven photonic cavities coupled to quantum materials~\cite{curtis2019cavity,schlawin2019cavity,chakraborty2020non,sentef2018cavity,thomas2019exploring}.
The complexity in modelling the physical principles  behind these platforms result from the necessity to  combine materials science together with an understanding of the role of driven photonic and/or phononic degrees of freedom in many-particle physics~\cite{laussy2010exciton,cotlect2016superconductivity,smolka2014cavity,mazza2019superradiant,andolina2019cavity,kiffner2019manipulating,latini2019cavity,li2020electromagnetic,ashida2020quantum,gao2020photoinduced,garcia2021manipulating,hubener2021engineering,li2020manipulating,lenk2020collective,chiocchetta2021cavity,ashida2021cavity,raines2020cavity,gao2021higgs}.  
It would be therefore desirable to provide an emulator of superconductivity which, although it may simplify the degrees of freedom involved, could shed light on complementary mechanisms for  non-equilibrium enhancement of superconducting order. 
This could then be used as a stepping stone towards richer and more intricate scenarios. 

\begin{figure}[h!]
    \centering
    \hspace*{-0.5cm}
    \includegraphics[height=0.7\columnwidth]{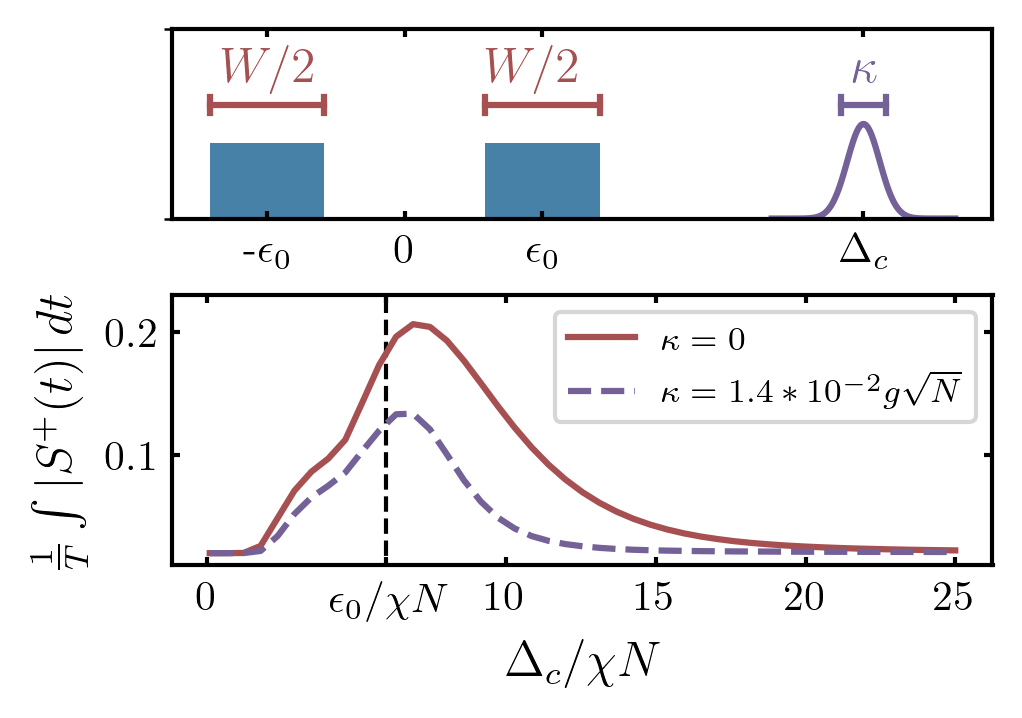}
    \caption{
        \textit{Top panel:} A simple schematic of the distribution of atomic levels (see text) and the spectral response of the cavity with detuning $\Delta_c$ and linewidth $\kappa$.
        \textit{Bottom panel:} Time average of the phase coherence $S^{+}=\frac{1}{N}\sum_{i}\left<\hat{\sigma}^+_i\right>$ as a function of the cavity detuning $\Delta_c/(\chi N)$, for large disorder $W/(\chi N)=8$.
        In the adiabatic limit, $\Delta_c/(\chi N)\rightarrow\infty$, the simulator shows vanishing coherence, $S^+(t)\rightarrow 0$, while coherence is maximum when the photon is at resonance with the mean atomic transverse field $\Delta_c\approx \epsilon_0=6 \chi N$~(marked by a black dashed line).
        To make use of an integrability analysis we assume an ideal cavity with $\kappa=0$ for most of the paper, and then confirm that a realistic cavity linewidth, $\kappa/(g\sqrt{N})=1.4\times 10^{-2}$, does not significantly modify the resonance phenomenon. 
    }
    \label{fig:main}
\end{figure}

Such an emulator has been proposed in AMO physics for quantum simulation of   archetypal s-wave superconductors (for charged particles) or s-wave superfluids (for neutral particles)~\cite{smale2019observation,lewis-swan2021} . 
In these works the dynamics of the superfluid phase coherence, directly related to the Meisner  and  Anderson-Higgs mechanisms in superconductors~\cite{annett2004}, can be   studied by monitoring the dynamics of the atomic phase coherence.
In the QED simulators considered so far, the cavity must be far detuned from atomic frequencies so that photonic degrees of freedom can be integrated out~\cite{agarwal1997,damanet2019,PhysRevLett.126.133603,muniz2020,norcia2018b,palacino2020,davis2019,periwal2021programmable,PhysRevX.8.011002,PhysRevX.11.021048,seetharam2021correlation,klinder2015dynamical,mivehvar2021cavity,baumann2010dicke,brennecke2007cavity,fogarty2015,habibian2013,himbert2019,Keller_2018,Keller_2017,schutz2016,PhysRevLett.116.153002} and so an effective matter-only s-wave model of superconductivity is sufficient to describe the dynamics.
In such a limit the cavity only contains virtual photons, and their primary purpose is to mediate  pairing interactions.

In this Letter, we investigate the effect of real photons on the phase coherence when the cavity detuning to the atomic transition is reduced.
In this limit, the single channel s-wave BCS Hamiltonian is no longer an accurate description, and instead the atoms and cavity field simulates the two channel model of the BCS-BEC crossover~\cite{yuzbashyan2015,annett2004}. 
In this model, the effect of reducing photon detuning on the dynamics are non-trivial because, on one hand, reducing the detuning yields a stronger mediated interaction strength, while on the other hand reducing the detuning leads to retarded photon dynamics where an instantaneous interaction is no longer valid.
Here, we find that even when the change in interaction strength is accounted for, the retarded photon dynamics can maintain phase coherence better than the instantaneous interaction.
This is demonstrated in Fig.~\ref{fig:main}, where we show that upon reducing the photon detuning phase coherence increases until resonance, below which the diabatic (small detuning) limit takes over and phase coherence is lost.
While these results are mostly obtained by a classical integrability analysis~\cite{dukelsky2004,richardson2002,PhysRevLett.96.230403,barankov2004collective,yuzbashyan2005,yuzbashyan2005a,yuzbashyan2006,yuzbashyan2015}, we also find via numerical simulation that the phenomenon is robust to the non-integrable effects caused by inhomogeneous couplings and photon loss which are typically present in realistic cavity QED settings. 

\textbf{Simulation of Superfluid Phase Coherence.} 
We consider the simulation of the two-channel model for the BCS-BEC crossover observed in ultracold fermion experiments~\cite{yuzbashyan2015,annett2004}.
The model involves fermions (with creation operator $\hat{f}^{\dagger}_{\textbf{k},s}$ with momentum vector $\textbf{k}$ and spin $s$) that can form Cooper pairs on the BCS side of the crossover or, bind into diatomic bosonic molecules at zero center of mass momentum~(with creation operator $\hat{d}^{\dagger}$) on the BEC side of the crossover.   Neglecting finite momentum molecular bosons, the dynamics are characterized by the Hamiltonian:
\begin{eqnarray*}
    H_f/\hbar&=& \sum_{\textbf{k},s}\epsilon_{\left|\textbf{k}\right|} \hat{f}^{\dagger}_{\textbf{k},s}\hat{f}_{\textbf{k},s} 
     +g \sum_{\textbf{k}}\hat{f}^{\dagger}_{\textbf{k},\uparrow}\hat{f}^{\dagger}_{-\textbf{k},\downarrow}\hat{d} + h.c.+ \Delta_c \hat{d}^\dagger \hat{d},
\end{eqnarray*}
where $\hat{d}$ is the mean molecular field, $\Delta_c$ is the molecular binding energy, and $g$ is the coupling strength between fermions and molecules.
When the fermions condense into a superfluid on the BCS side of the crossover, they mostly form Cooper pairs~\cite{annett2004} quantified by the complex pair amplitudes $\rho_{\textbf{k}}=\langle\hat{f}^{\dagger}_{\textbf{k},\uparrow}\hat{f}^{\dagger}_{-\textbf{k},\downarrow}\rangle$.
In this Letter, we focus on the dynamics of the superfluid s-wave phase coherence $S^+=\frac{1}{N}\sum_{\textbf{k}}\rho_{\textbf{k}}$, which quantifies the phase coherence between Cooper pairs with different pairing wave vector $\textbf{k}$.

Similar to Ref.~\cite{lewis-swan2021}, the Cooper pairs can be simulated by a collection of two level atoms (described by Pauli operators $\hat{\sigma}^{+}_i$ and $\hat{\sigma}^{z}_i$) via the Anderson pseudospin mapping~\cite{PhysRevLett.96.230403,barankov2004collective,yuzbashyan2015}:
\begin{eqnarray}
    \resizebox{.8\hsize}{!}{	$   \hat{\sigma}^{+}_{i}\rightarrow \hat{f}^{\dagger}_{\textbf{k}_i,\uparrow}\hat{f}^{\dagger}_{-\textbf{k}_i,\downarrow}, \quad \hat{\sigma}^{z}_{i}\rightarrow \hat{f}^{\dagger}_{\textbf{k}_i,\uparrow}\hat{f}_{\textbf{k}_i,\uparrow}+\hat{f}^{\dagger}_{-\textbf{k}_i,\downarrow}\hat{f}_{-\textbf{k}_i,\downarrow}-1$}.
\end{eqnarray}
where each atom $i$ simulates a pair of fermion momentum modes $i \rightarrow \pm \textbf{k}_i$.
The above Hamiltonian can then be simulated by a cavity QED system similar to the experiments described in references~\cite{norcia2018b,Norcia2016,muniz2020,davis2019}, 
in which the internal levels of $2N$ atoms are encoded in long lived electronic states, e.g the ${}^1S_0$-${}^3P_1$ states of $^{88}$Sr atoms. The atoms are trapped in an optical lattice and are allowed to interact with a single cavity mode (described by a photon annihilation operator $\hat{a}$  simulating the molecular field, $\hat{a}\rightarrow\hat{d}$).
Such a system is modeled by the Hamiltonian~\cite{norcia2018b,muniz2020,lewis-swan2021}:
\begin{eqnarray}\label{eq:ham}
    H/\hbar=\sum_{i=1}^{2N} \epsilon_i \hat{\sigma}^z_i +\sum_{i=1}^{2N} g_i (\hat{\sigma}^{\dagger}_i \hat{a}+ h.c.) + \Delta_c \hat{a}^{\dagger}\hat{a},
\end{eqnarray}
where $\Delta_c$ is the detuning of the cavity from the mean atomic frequency, $2g_i$ is the single-photon Rabi frequency, and $\epsilon_i$ is an inhomogeneous effective transverse field.
Simulation of $H_f$ by the cavity QED system occurs for homogenous light-matter coupling $g_i=g$ and for a probability distribution, $p(\epsilon_i)$, of the inhomogeneous field, $\epsilon_i$, that is designed to match the density of states for the fermion model. 
We choose the density of states as $p(\epsilon_i)=\left[B(W/2,\epsilon_0/2,\epsilon_i)+B(W/2,-\epsilon_0/2,\epsilon_i)\right]/2$, where $B(\alpha,x_0,x)$ is a box distribution with mean $x_0$ and width $\alpha$~(see Fig.~\ref{fig:main}). {Similar to Ref.~\cite{lewis-swan2021}, such a bimodal distribution is chosen to ensure the possibility of persistent oscillations of the phase coherence (see below) in the $W=0$ limit.}
A possible band structure reproducing this density of states and the superfluidity that would occur in the traditional thermal equilibrium setting is discussed in Ref.~\footnote{See SM for band structure and equilibrium phase diagram}. 

At large detuning, $\Delta_c\gg g\sqrt{N}$ and $\Delta_c\gg \epsilon_0+W/2$, the cavity field mediates spin-exchange interactions and an effective spin model  can be derived which maps into a one channel BCS model as discussed in Ref.~\cite{lewis-swan2021}.
In this limit, an adiabatic approximation~\cite{norcia2018b,muniz2020,lewis-swan2021} assumes the state of the light field is in instantaneous equilibrium such that $\left<\hat{a}_{eq}^{\dagger}(t)\right>= -\frac{gN}{\Delta_c}S^+$, where $S^+=\frac{1}{N}\sum_i \left<\sigma_i^+\right>=\frac{1}{N}\sum_{\textbf{k}}\rho_\textbf{k}$ is both the atomic phase coherence and the simulated superfluid phase coherence.
Thus, in the large detuning limit, the photon directly measures the phase coherence $S^+$.
Inserting $\left<\hat{a}_{eq}^{\dagger}(t)\right>$ back into Eq.~\ref{eq:ham} and taking homogenous couplings, one finds a mediated interaction $-\chi\sum_{ii'}\hat{\sigma}^{+}_i\hat{\sigma}^{-}_{i'}$ with interaction strength $\chi=g^2/\Delta_c$ {and sign which favors effective Cooper pair formation at low temperatures and positive detuning, $\Delta_c$}.
In this work we will study the dynamics when the photon detuning, $\Delta_c$, is decreased and the adiabatic approximation is no longer valid.
One complication to this limit is that when the photon detuning is decreased, the interaction strength $\chi$ increases.
To isolate this effect we imagine that the experiment simultaneously increases the strength of the atomic energies as the photon detuning is decreased such that $\epsilon_0/\chi N$ and $W/\chi N$ are held constant.

\textbf{Dynamical Phases from classical integrability.}
To study the dynamics of this system, we make a mean field approximation (i.e. $\langle\hat{O}_1(t)\hat{O}_2(t)\rangle=\langle\hat{O}_1(t)\rangle\langle\hat{O}_2(t)\rangle$ and adopt the notation:~$\langle\hat{O}_1\rangle\equiv O_1$) which is expected to work up to time scales $O(1/(\chi N))$~\cite{kirton2017,kirton2018,kirton2019}.
The resulting classical dynamics of the Hamiltonian in Eq.~\ref{eq:ham} show Richardson Gaudin integrability~\cite{dukelsky2004,richardson2002,PhysRevLett.96.230403,barankov2004collective,yuzbashyan2005,yuzbashyan2005a,yuzbashyan2006,yuzbashyan2015,RGIntreview,RICHARDSON1964,gaudin1976} in the homogenous limit, $g_k=g$ .
The so called Lax integrability analysis~\cite{dukelsky2004,richardson2002,PhysRevLett.96.230403,barankov2004collective,yuzbashyan2005,yuzbashyan2005a,yuzbashyan2006,yuzbashyan2015} is then used to study the integrable tori of the classical mean field Hamiltonian corresponding to Eq.~\ref{eq:ham} and to construct a dynamical phase diagram~\cite{lewis-swan2021,yuzbashyan2015} characterizing the collective modes. 
This is done by studying the conserved quantities to identify a minimum number, $M$, of collective degrees of freedom (DOF) required to effectively reproduce the dynamics of collective variables at long times~\footnote{see SM for details on Lax Analysis}.

\begin{figure}[h!]
    \centering
    \hspace*{-0.5cm}\includegraphics[width=\columnwidth]{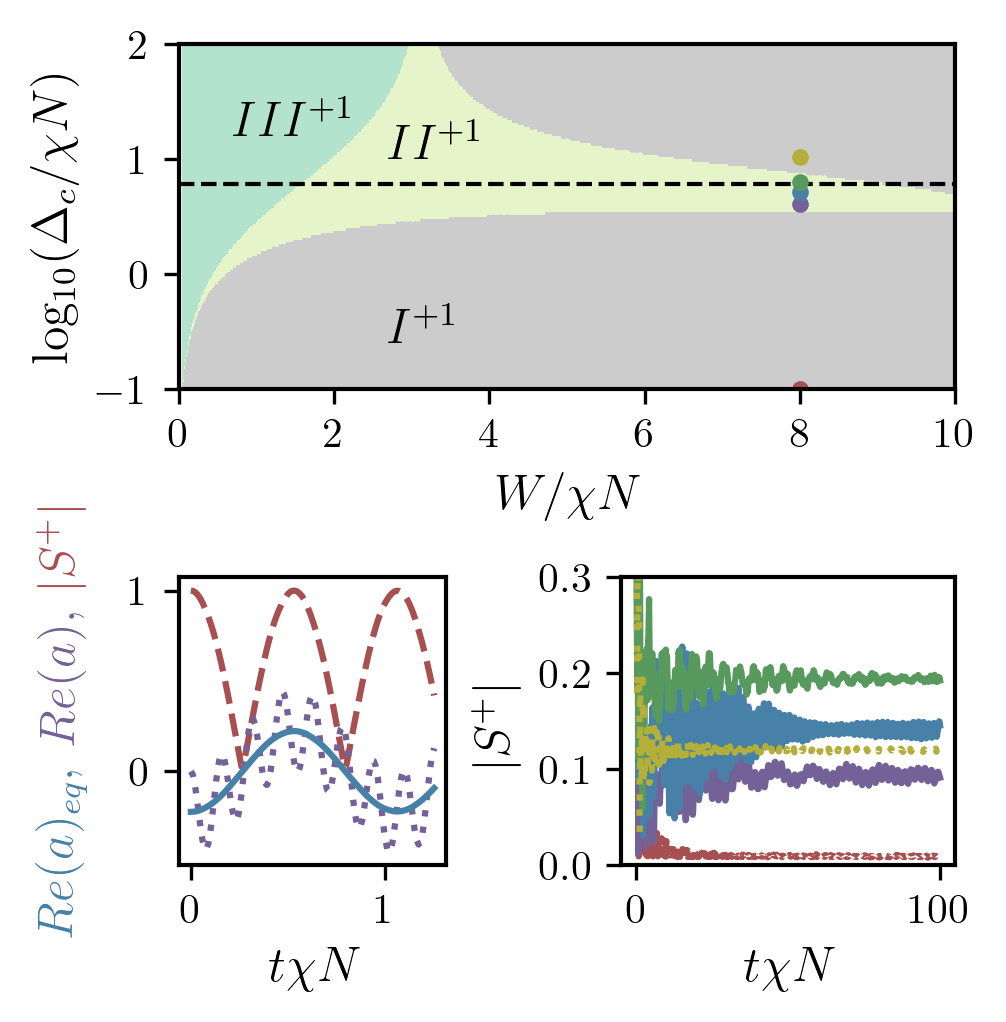}
    \caption{ 
        The top panel shows the dynamical phase diagram as a function of the cavity detuning and atomic disorder.
        The approximate resonance condition, $\Delta_c\approx\epsilon_0=6\chi N$, is marked by a black dashed line.
        The bottom left panel shows dynamics characteristic of phase $III^{+1}$ (specifically $W=0$ and $\Delta/(\chi N)=40$). For these parameters, the adiabatic approximation correctly predicts the dynamics of the matter, but misses the extra oscillation around $a_{eq}(t)$ predicted by the Lax analysis.
        The bottom right panel shows the dynamics of $\left|S^+(t)\right|$ for $W/(\chi N)=8$ and different values of $\Delta_c/(\chi N)$ marked in the top panel.
        For $\Delta_c/(\chi N)=0.1$ (red) and $10.3$ (yellow), $\left|S^+(t)\right|$ evolves to a constant steady state characteristic of phase $I^{+1}$, while for the remaining values of $\Delta_c/(\chi N)$ the dynamics have persistent oscillations characteristic of phase $II^{+1}$. In this figure, the initial state has $\left<\hat{\sigma}^x_i\right>=1$ and $\left<a\right>=0$.
    }
    \label{fig:dyn}
\end{figure}

The dynamical phases are then classified by the required number of collective DOF and the dynamics of the phase coherence $S^+$.
First, we consider the resulting collective modes for a quench starting from an initial state with all spins polarized in the $\hat{x}$ direction, $\left<\hat{\sigma}^x_i\right>=1$, and the cavity in the vacuum, $\left<a\right>=0$. 
In the spin-only model, three phases are found~\cite{yuzbashyan2015,lewis-swan2021} with at most $ M=2$. In contrast, we identify a fourth phase with $M=3$ upon introducing the photon away from adiabatic elimination.
The three phases in the adiabatic limit, $\Delta_c\rightarrow\infty$, are (for fixed $\chi N$ and $\epsilon_0>\chi N$):
\begin{itemize}
    \item \textit{Phase I}  ($M=0$): At large disorder, all phase coherence is lost, and the simulated superfluid enters a normal state: $S^+(t)\rightarrow 0$. 
    \item \textit{Phase II} ($M=1$): Transition to this phase occurs as disorder is reduced, and involves only one effective degree of freedom ($M=1$). In this phase, the magnitude of the phase coherence, $\left|S^+(t)\right|$, is constant at late time, and the collective mode corresponds to precession of the phase of $S^{+}$: $S^+(t)\rightarrow\left|S^{+}\right| e^{i\mu t}$.
    \item \textit{Phase III} ($M=2$): This phase occurs at even smaller inhomogeneous atomic broadening, and has $M=2$ DOF.
The collective mode shows persistent oscillations in $\left|S^+(t)\right|$ as shown in the lower panel of Fig.~\ref{fig:dyn}.
\end{itemize}
In this adiabatic limit, the critical disorder strengths between phase I and II, and II and III, depends non-trivially on $\epsilon_0$, but are on the order of the interaction strength $\chi N$.

At finite detuning, $\Delta_c$, the photon becomes another DOF in the collective oscillations of these three phases, and to distinguish the phases of the full model we will write them with a ``$+1$'' superscript.
The phases $I^{+1}$ and $II^{+1}$ show the same qualitative dynamics of $S^+(t)$ as the phases $II$ and $III$ respectively, while a new phase $III^{+1}$ is defined by aperiodic oscillations of $\left|S^+(t)\right|$ and requires $M=3$ collective DOF (two macroscopically coherent spins and a photon). 
At large but finite $\Delta_c$, the new phase $III^{+1}$ involves the photon performing fast oscillations around $a_{eq}(t)$, the slowly evolving equilibrium value given by adiabatic elimination (see Fig.~\ref{fig:dyn} for an example).
In this limit, the aperiodic contribution to the oscillations of $S^+$ becomes small smoothly as function of $\Delta_c$, and thus in the large detuning limit phase $III^{+1}$ approximates phase $III$.
This limiting behavior is the same for phases $I^{+1}$ and $II^{+1}$ which, for large detuning, approximate phase $I$ and $II$ respectively.

Upon reducing the detuning, a rich dynamical phase diagram emerges as shown in the upper panel of Fig~\ref{fig:dyn}.
In the $W=0$ limit, there is only phase $III^{+1}$, while, at finite $W$,  the cavity field has a broad impact on the dynamical phase diagram.
In the diabatic limit, $\Delta_c/\chi N\ll 1$, the dynamics are much more sensitive to the inhomogeneities due to an inability of the cavity to mediate an effective interaction, and the transition to phase $I^{+1}$ occurs at much smaller disorder in comparison to the large detuning limit.
We also find a region at large disorder, $W>4\chi N$, where phase $II^{+1}$ occurs when $\Delta_c\sim \epsilon_0$ which suggests phase coherence can be enhanced by setting the detuning on resonance with the atoms that have atomic energies close to $\epsilon_0$.

\textbf{Mechanism of resonant phase coherence.} {
The enhancement of phase coherence is confirmed in Fig.~\ref{fig:main}, and we explain the formation of this resonance by first considering finite but large detuning, such that $1/\Delta_c$ is still the fastest timescale. In this limit the dynamics are in Phase $I^{+1}$ and the enhancement of phase coherence is very weak at long times, but the following simple picture holds.
First, on a timescale of $1/\Delta_c$, the initial polarization of the spins drive the photon into an excited state oscillating around a non-zero $a_{eq}(t=0)=-Ng/\Delta_c$.
Then, on a timescale of $1/W\gtrsim1/\Delta_c$, the spins mostly dephase and $a_{eq}(t\rightarrow\infty) \approx 0$.
Once the spins mostly depolarize to their steady state, the photon remains oscillating around a small equilibrium, $a(t)\approx Ae^{i\mu t}$.
The Lax analysis (see~\cite{yuzbashyan2015} and~\cite{Note1}) yields expressions for the frequency and amplitude of these small oscillations which have a simple analytical form when $\Delta_c>W>\epsilon_0$: $\mu=\Delta_c$ and $A=\chi N/g=\sqrt{N}\sqrt{\chi N/\Delta_c}$. 

From the perspective of the matter, the photon is effectively an external drive that pumps a small fraction of the spins into a coherent steady state.
In the frame of reference of the photon (the effective external drive), the dynamics of each spin is fully described by a constant magnetic field, $\vec{h}=(\epsilon_i-\mu) \hat{z}+gA\hat{x}$, and we can solve for the steady state as:
\begin{eqnarray}\label{eq:res}
    S^+=\frac{1}{N}\sum_i\frac{gA}{\sqrt{(\epsilon_i-\mu)^2+\left|gA\right|^2}}.
\end{eqnarray}
Since $\mu\sim\Delta_c$, this expression correctly predicts the loss of coherence, $S^+$, in the adiabatic limit shown in Fig.~\ref{fig:main}.

{Further away from the adiabatic limit, the separation of time scales, $1/W\gtrsim 1/\Delta_c$, that yields the simple picture above is no longer valid.
Regardless, the Lax analysis still produces the same expression, Eq.~\ref{eq:res}, for the phase coherence in phase $I^{+1}$ but now with a different $A$ and $\mu$ that must be numerically determined by solving for the roots of a Lax vector (see~\cite{yuzbashyan2015} and SM).
Since $\mu$ gives the precession frequency of the photon, it is expected to be close to the detuning $\mu\approx\Delta_c$ and this is what we find numerically. }
Eq.~\ref{eq:res} therefore predicts the atom at site $i$ will be in resonance when $\Delta_c=\epsilon_i$.
The coherence is then maximally enhanced when most spins are driven close to resonance and occurs when the drive, $\Delta_c$, is at the center of the band of atomic frequencies $\Delta_c\approx\epsilon_0$.
This approximation is confirmed by the peak in {coherence} shown in Fig~\ref{fig:main}. 

Although  Eq.~\ref{eq:res} provides an intuitive picture, simliar to a single particle resonance, when the system is in phase $I^{+1}$, the relevant enhancement of coherence at the resonance happens in phase $II^{+1}$ where the cavity field  and atomic coherence must both be treated as dynamical variables. 
As shown in Fig.~\ref{fig:dyn}, their dynamics in this regime show coupled nonlinear oscillations~\cite{yuzbashyan2015}. 

\footnotetext{See SM for details on mean field approximation for Lindblad Evolution}
\begin{figure}[]
    \centering
    \hspace*{-0.5cm}\includegraphics[width=\columnwidth]{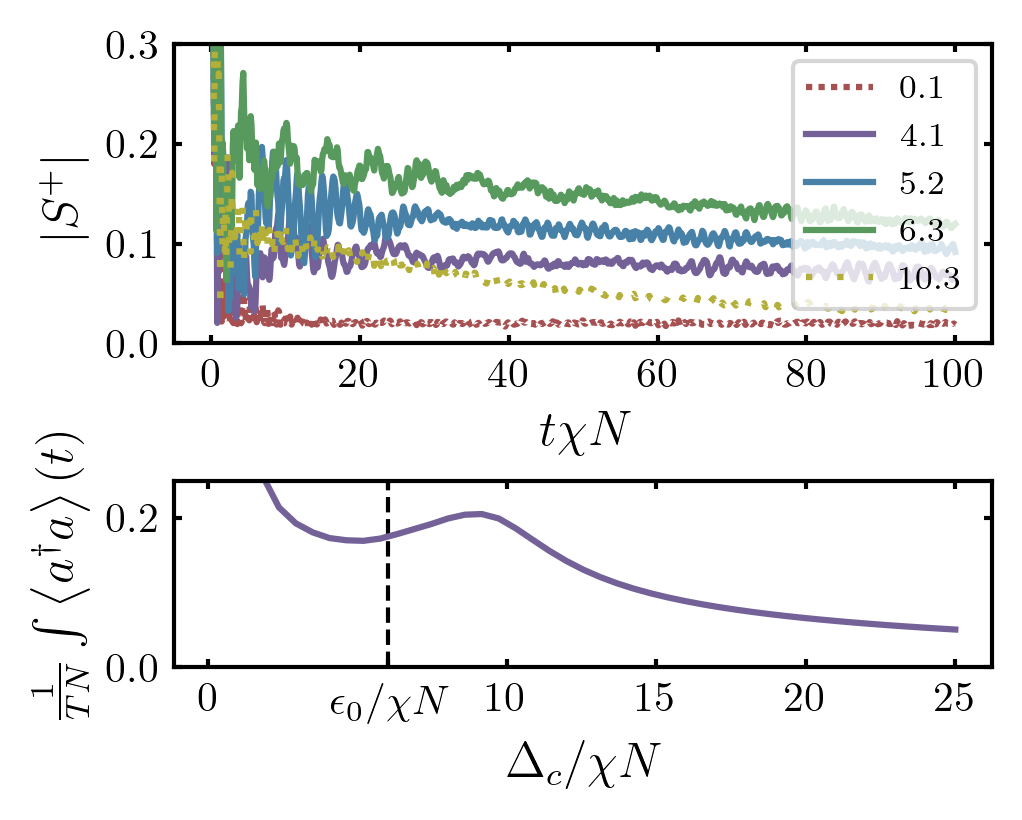}
    \caption{
        The top panel shows the same dynamics for $\left|S^+(t)\right|$ as in Fig.~\ref{fig:dyn}, but with $\kappa/(2\pi)=150~{\rm kHz}$ and {$g/(2\pi)=10.8~{\rm kHz}$} as in the experiments of~\cite{muniz2020,norcia2018b}.
        The bottom panel is computed  assuming inhomogeneous couplings and an initial state  prepared by a coherent drive through the cavity.
        Even though inhomogeneities  and cavity losses reduce the coherence, we can observe a signature of the resonance as a minimum~\cite{Note4} in the photon density. 
        Note that in contrast to Fig.~\ref{fig:dyn} and Fig.~\ref{fig:main}, the  initial state depends on the initial number of photons driven  into the cavity which scales as $1/\Delta_c$.
        Both figures were obtained by numerical simulation of the Lindblad equations of motion at mean-field~\cite{Note3}.
        In the top panel, the decay rate of $\left|S^+(t)\right|$ is constant with $\Delta_c/\chi N$ and proportional to $1/\kappa$, but appears to increase with $\Delta_c/\chi N$ in the figure because the unit for time, $1/\chi N$, decreases with $\Delta_c$ when $g$ is fixed.
    }
    \label{fig:exp}
\end{figure}

\textbf{Experimental Realization.} In the experiments of Refs.~\cite{muniz2020,norcia2018b}  an optical lattice is used to trap Sr atoms, featuring a long-lived electronic clock transition with atomic decay rate of $\gamma$. 
The optical lattice is placed inside a standing wave optical cavity  with linewidth {$\kappa$}.
While both $\gamma$ and $\kappa$ destroy phase coherence at long times, we find that the effect of resonant phase coherence is still observable on times $O(1/\kappa)$ provided we operate at large collective cooperativity ($N g^2\gg \kappa \gamma$).
Given that for long-lived Sr atoms, $\kappa\gg \gamma$ we neglect atomic decay.
Fig.~\ref{fig:main} shows the dependence of $\left|S^+(t)\right|$ on $\Delta_c$, and demonstrates that the resonant enhancement can be maintained even with cavity loss.
Furthermore, Fig.~\ref{fig:exp} depicts how the dynamics in Fig.~\ref{fig:dyn} simply features a slow decay for moderate $\kappa$. 

The experiments in~\cite{muniz2020,norcia2018b} also have inhomogeneous couplings {$g_i=g\cos(k_0 a_0 i)$} with $k_0 a_0=3.7$ due to an incommensurability between the optical lattice spacing, $a_0$, and the cavity wavelength, $2\pi/k_0$.
The inhomogeneous couplings will disrupt the effect discussed in this work if we start in a homogenous state, since the couplings will no longer excite the photon.
However, as long as the initial state is generated by coherently driving the optical cavity, inhomogeneities do not play a detrimental effect. In this case, the initial state involves all spins aligned with the inhomogeneities $\text{sgn}(\sigma^x_i)=\text{sgn}(g_i)$ such that cavity will be coherently pumped by the atoms.
The resulting simulations show signature of resonant phase coherence as a minimum~\footnote{See SM for the relation between resonance and minimum of photon density} of the time averaged photon density shown in Fig.~\ref{fig:exp}.
Note that both dissipation and inhomogeneities break Lax integrability.

\textbf{Conclusion.}
Our work demonstrates that dynamical fluctuations of a mediating field can produce enhancement of phase coherence in cavity QED simulators of superconductivity and superfluidity. 
The generality of our result based on a resonance argument, would suggest a natural extension to a broad variety of platforms such as trapped ions or quantum optics in waveguides, both of which serve as tunable simulators of non-equilibrium quantum many body physics, employing mediating photons or phonons~\cite{ions2019,RevModPhys.90.031002}. 
It also suggests a promising direction for cavity enhanced superconductivity in real materials.
Such a possibility requires extending the phenomenon to charged superfluids in which the light-matter couplings are structurally different from the atom-molecule couplings of Eq.~\ref{fig:main}. 
Our results offer the possibility of studying novel regimes of enhanced cooperative light-matter, and hint that  quantum many-body optics with active light and matter degrees of freedom has the potential to become a blossoming area of quantum simulation in the near future.

 \emph{Acknowledgments-- }    
This work has been funded by the Deutsche Forschungsgemeinschaft (DFG, German Research Foundation) - TRR 288 - 422213477 (project B09), TRR 306 QuCoLiMa (”Quantum Cooperativity of Light and Matter”), Project-ID 429529648 (project D04) and in part by the National Science Foundation under Grant No. NSF PHY-1748958 (KITP program ‘Non-Equilibrium Universality: From Classical to Quantum and Back’). 
A.  M.  R.  acknowledges  ARO  (Army Research  Office)  under  the  Grant  No.    W911NF-19-1-0210, and the  National Science Foundation under the Grants No.  NSF  PHY1820885. Both A.M.R and J.K.T   acknowledge NSF JILA-PFC PHY-1734006, QLCI-OMA -2016244,  the U.S. Department of Energy, Office of Science, National Quantum Information Science Research Centers Quantum Systems Accelerator, and NIST.

\bibliography{refs.bib}

 \onecolumngrid
 
\appendix

\section{Lax analysis for quench dynamics}
In the main text, we studied the dynamics of an initial state with $\left<\sigma^x_i\right>=1$ and with the photon in an empty cavity state. Here, we consider a more general state used in~\cite{lewis-swan2021} where the spins $1\dots N$ point in the $\hat{x}\cos(\Delta \phi_{0}/2)+\hat{y}\sin(\Delta \phi_{0}/2)$ direction, while the spins $N+1\dots2N$ point in the $\hat{x}\cos(\Delta \phi_{0}/2)-\hat{y}\sin(\Delta \phi_{0}/2)$ direction, and the photon is either in the vacuum, $a(t=0)=0$ or at equilibrium with matter, $a(t=0)=a_{eq}(0)=-\frac{1}{\Delta_c}\sum_i g_i \sigma^{-}_i(t=0)$.

We now compute the lax vector for such a state:
\begin{eqnarray}
    \vec{L}(u,t)=\sum_i \frac{\vec{\sigma}_i(t)}{u-\epsilon_i}+\hat{z}\frac{2}{g^2}(u-\Delta_c/2)-\frac{2\vec{\Delta}(t)}{g^2}
\end{eqnarray}
Since the initial state is uniform for spins $1\dots N$ and for $N+1\dots2N$, the sums over $i$ splits into two sums which can be approximated by disorder average:
\begin{eqnarray}
    \sum_{i}\frac{1}{u-\epsilon_i}\approx N\int \frac{p(\epsilon)}{u-\epsilon}d\epsilon
\end{eqnarray}
where $p(\epsilon)$ is the disorder distribution for transverse fields: $\epsilon_i=\epsilon/2+x_i$ for $i\in(1\dots N)$ and $\epsilon_i=-\epsilon/2+x$ for $i\in(N+1\dots 2N)$, and $x_i$ are drawn from a uniform distribution with zero mean and width $W/2$.
The integral results in a logarithm whose branch cut must be chosen to match the continuum of poles that develop when taking the $N\rightarrow\infty$ limit of the summation\cite{yuzbashyan2005,yuzbashyan2005a,yuzbashyan2006}.
This results in:
\begin{eqnarray}
    \lim_{N\rightarrow \infty}\frac{\chi}{2} \sum_i \frac{\sigma^{y}_i(t=0)}{u-\epsilon_i}&=L^y_{l}(u)=&\frac{\chi N}{W} \sin(\frac{\Delta \phi_0}{2})\ln\left[\frac{(4u)^2-(W-2\epsilon_0)^2}{(4u)^2-(W+2\epsilon_0)^2}\right] \\ \nonumber
    \lim_{N\rightarrow \infty}\frac{\chi}{2} \sum_i \frac{\sigma^{x}_i(t=0)}{u-\epsilon_i}&=L^x_{l}(u)=&2\frac{\chi N}{W}\cos(\Delta \phi/2) \left[\text{Arctanh}(\frac{W-2\epsilon_0}{4u})+\text{Arctanh}(\frac{W+2\epsilon_0}{4u})\right],
\end{eqnarray}
where the branch cut for $\text{ln}(x)$ is $Re(x)\in[0,-\infty)$, and the branch cut for $\text{ArchTanh}(x)$ is $Re(x)\in[-1,-\infty)$ and $Re(x)\in[1,\infty)$.
The square lax vector is therefore computed as:
\begin{eqnarray}
    \frac{\chi}{2}\vec{L}^2=\left(\chi\vec{L}^x_{l\perp}(u)-s\frac{\chi N}{\Delta_c}\cos(\Delta\phi/2)\right)^2+\chi\vec{L}^y_{l\perp}(u)^2+\left(1-\frac{2u}{\Delta_c}\right)^2
\end{eqnarray}
where $s=0$ if the cavity field is $a(t)=0$ at $t=0$, or $s=1$ if the cavity field is in equilibrium with matter $a(t=0)=a_{eq}(t=0)$ at $t=0$.

In the main text, we used the numerical search described below to study the dynamical phases as a function of $\Delta_c/\chi N$ and $W/\chi N$, for fixed $\epsilon_0/\chi N=6$, $\Delta\phi_0=0$ and   photon starting in the vacuum. 
Such a phase diagram does not qualitatively change by increasing $\Delta\phi$, but if the cavity field is in equilibrium with matter $a(t=0)=a_{eq}(t=0)$ at $t=0$, and $\Delta\phi_{0}=\pi$ we see larger region for phase $II^{+1}$ (see Fig~\ref{fig:SMphaseDiagram}).

\begin{figure}[h!]
    \centering
    \includegraphics[height=2.5in]{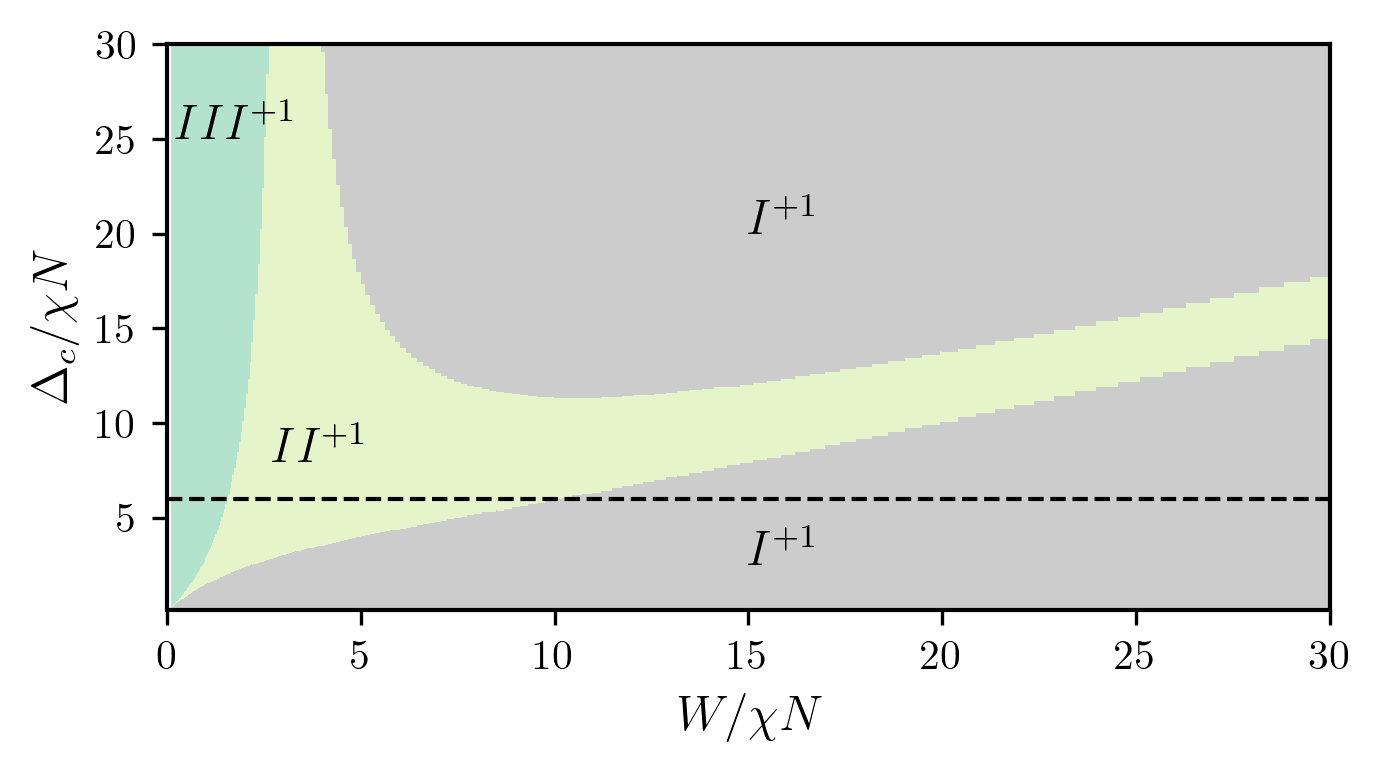}\\
    \caption{Phase diagram as in main text but for an initial state with $\Delta\phi_0=\pi$ and $a(0)=a_{eq}(0)$. Interestingly, for large $W$, the region of Phase $II^{+1}$ shifts to higher frequencies. While this region is in phase $II^{+1}$ where the amplitude $\left|S^+\right|$ oscillates, the oscillations and over all phase coherence are relatively small at large $W$.  Similarly, the Lax analysis of phase $I^{+1}$ still predicts a resonance in phase coherence $\left|S^+\right|$ at $\Delta_c=\epsilon_0$ (marked by a black dashed line); we have confirmed this numerically.}
    \label{fig:SMphaseDiagram}
\end{figure}

\subsection{Numerical Search for the roots}
The dynamical phases are characterized by the number of roots of $\vec{L}^2(u)=0$ as discussed in the main text.
To find these roots for a range of $W/\chi N$ and $\Delta_c/\chi N$ as in the main text, we employ the NLsolve Julia library.
The algorithm   requires an initial seed for the roots and then uses the steepest decent to find the numerical value.
Therefore, to find all the unique roots $\vec{L}^2(u)=0$ and thereby correctly identify the dynamical phase we perform the following process.
First, we plot the complex magnitude of $\vec{L}^2(u)$ and identify an approximate location of the 6 roots from the plot.
Then we use NLsolve function to find a precise location of the root with a relative and absolute numerical error tolerance of $10^{-4}$.
For the first root, this is done for $W=0$, a specific initial state, a large value of $\Delta_c/\chi N$, and the remaining Hamiltonian parameters fixed.

Since the location of the roots are smoothly connected to each other as $\Delta_c/\chi N$ is smoothly decreased, we next find the location of roots for $W/\chi N=0$ and different values of $\Delta_c/\chi N$. 
This is done by starting with roots found for $W=0$ and the large value of $\Delta_c/\chi N$ and then finding the roots sequentially decreasing $\Delta_c/\chi N$.
The seeds used for the NLsolve function are the roots obtained from the previous step in the sequence with slightly larger $\Delta_c/\chi N$.
Then, starting with the root found for a specific $\Delta_c/\chi N$ and $W/\chi N=0$, we find the roots for $W/\chi N>0$ performing a similar sequential process, but this time increasing $W/\chi N$ and keeping $\Delta_c/\chi N$ fixed.
After finding all the roots, we identify the unique ones, up to an error tolerance of $10^{-4}$, to identify which phase we are in.

A problem that can arise in this process is   two roots closely approach each other, as $W/\chi N$ is increased.
If this occurs, the seeds used for steepest decent can become identical to each other in the sequential update of seeds based on the previous roots.
In order to prevent such issue, we keep a list of fixed seeds that we also use when finding distinct roots at each point in the phase diagram.

\subsection{Analytic solution for phase $I^{+1}$ roots}
As discussed in the main text, the location of the two roots in phase $I^{+1}$ gives useful information on the dynamics of the photon. These roots can be found if
\begin{eqnarray}
    \Delta_c>W>\epsilon_{0},
\end{eqnarray}
when the cavity is initially empty, and when $\Delta\phi_0=0$.
In this limit, the $\text{ArchTanh}$ can be expanded, and the Lax squared vector yields:
\begin{eqnarray}
    \chi/2\vec{L}^2\approx(\frac{4\chi N}{W})^2(\frac{W}{4u})^2+(1-\frac{2u}{\Delta_c})^2+...
\end{eqnarray}
Defining $u=u'\Delta_c/2$, the condition $\vec{L}^2(u)=0$ becomes:
\begin{eqnarray*}
    \pm i\frac{2\chi N}{\Delta_c}\frac{1}{u'}=1-u'
\end{eqnarray*}
The quadratic equation solves as
\begin{eqnarray}
    u'=\frac{1\pm\sqrt{1\pm i8/\Delta_c'}}{2}.
\end{eqnarray}
When $\Delta_c'=\Delta_c/\chi/N$ is very large the small angle approximation yields $u'\approx1\pm i\frac{2}{\Delta_c'}$, or $ u=\frac{\Delta_c}{2}\pm i \chi N$.
The Lax analysis argues~\cite{yuzbashyan2015} that in Phase $I^{+1}$, the real and imaginary parts of this root pair gives the frequency, $\mu$, and amplitude, $A$, of the photon oscillations respectively: $a\approx Ae^{i\mu  t}$.
Therefore we find that when, $\Delta_c>W>\epsilon_0$ the photon frequency is given as $\mu=\Delta_c$ the photon amplitude as $A=\chi N/g=\sqrt{N}\sqrt{\chi N/\Delta_c}$. 
as in the main text.
This is confirmed in Fig.~\ref{fig:photon}.

\begin{figure}[t!]
    \centering
    \hspace*{-0.7cm}\includegraphics[width=0.5\columnwidth]{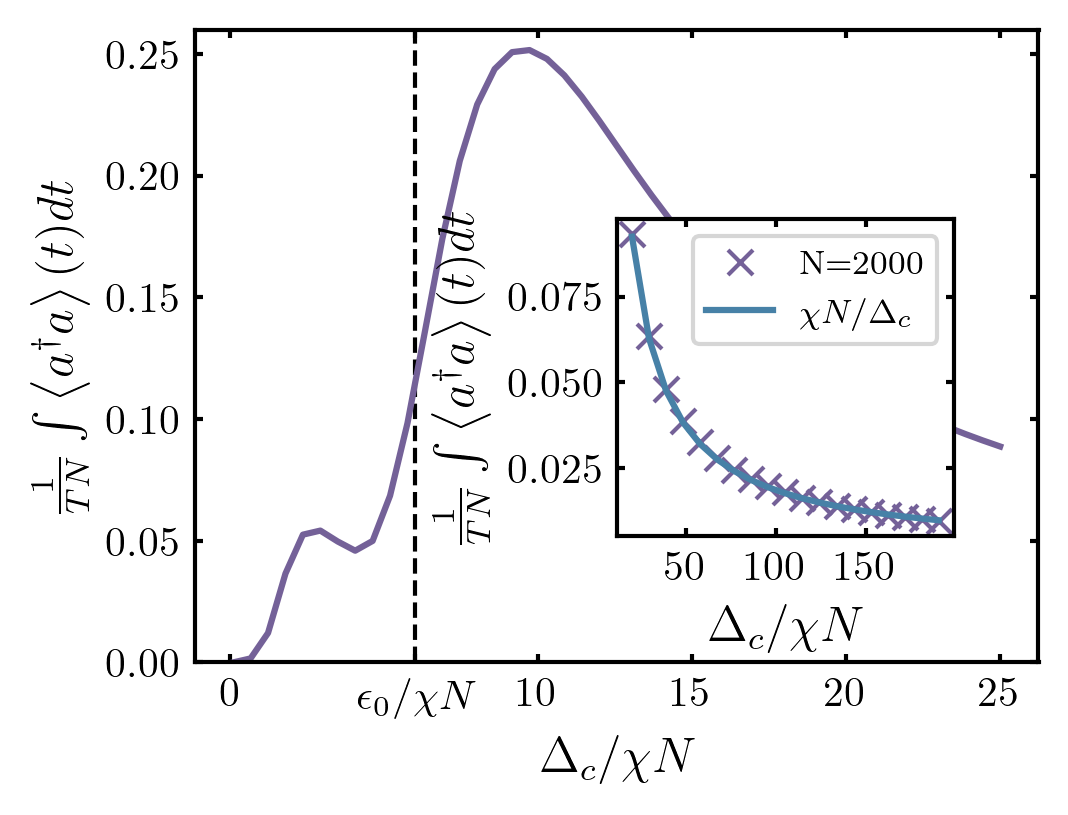}
    \caption{
        {This figure shows the effect of resonant {phase coherence} as monitored by the photons.} Since in the adiabatic limit  $a^\dagger\rightarrow-gNS^+/\Delta_c$, and $S^+$ vanishes at long times for the chosen conditions, it directly quantifies the extent to which adiabatic elimination fails.
        The inset shows numerical confirmation (marked by x's) of the analytic prediction for the photon amplitude $A^2=N^2\chi /\Delta_c$ (solid line). Such a prediction requires the approximation $\epsilon_0\ll W$, and therefore $\epsilon_0=\chi N \ll W=8\chi N$ in the inset, while in the main figure $\epsilon_0=6\chi N$. In both, the initial state has $\left<\hat{\sigma}^x_i\right>=1$ and $\left<a\right>=0$.
    }
    \label{fig:photon}
\end{figure}

\section{Minimum in photon occupation at resonance}

As seen in Fig.~\ref{fig:photon}, and in Fig.~3 of the main text, the photon density $\frac{1}{TN}\int_0^T dt \left<a^\dagger a\right>$ shows a minimum as a function of the detuning, $\Delta_c$, near resonance $\Delta_c=\epsilon_0$.
This phenomenon is explained in a similar way to the resonance maximum seen by the phase coherence $\left|S^+\right|$ and it is related to the conservation of total excitations: $S^z+a^{\dagger}a$.
In Fig.~\ref{fig:photon}, the initial state is polarized in the $\hat{x}$ direction and with no photons in the cavity such that $S^z+a^{\dagger}a=0$.
Therefore the photon density is fixed to the $J_z$ polarization of the spins: $a^{\dagger}a = - S^z$.
Using the same argument in the main text that, in phase $I^{+1}$, the photon simply acts as an external drive with $a=Ae^{i\mu t}$, we find that

\begin{eqnarray}
    a^{\dagger}a=-S^z=-\frac{1}{N}\sum_i\frac{(\epsilon_i-\mu)}{\sqrt{(\epsilon_i-\mu)^2+\left|gA\right|^2}}.
\end{eqnarray}
Therefore the $i^{th}$ spin contributes least to the total $S^z$ polarization (and photon amplitude) when it is in resonance with the cavity field $\epsilon_i=\mu\approx\Delta_c$.
This implies a minimum in $a^{\dagger}a$ when the spins on average are in resonance with the cavity field: $\epsilon_0=\Delta_c$.
Furthermore, a spin $i$ with frequency $\epsilon_i>\mu$ will have opposite $\hat{z}$ polarization than a spin $j$ with frequency $\epsilon_j<\mu$ resulting in overall reduction of their net contribution to the total spin polarization $S^z$.
This deconstructive interference between spins with different relative frequency $\epsilon_i-\mu$ will shift the minimum in $a^{\dagger}a$ to a smaller detuning  than $\Delta_c=\epsilon_0$ due to the contributions of spins in the negative band of frequencies: $\left|\epsilon_i+\epsilon_0\right|<W/2$.

The argument is similar for an initial state discussed in Fig.~3 of the main text, which also has $S^z(t=0)=0$ but with photons in the cavity, $\left|a(t=0)\right|=\left|a_0\right|>0$.
The only difference is the conservation condition gives an additional contribution to the photon density:
\begin{eqnarray}
    a^{\dagger}a=-J_z+a^{\dagger}_0a_0
\end{eqnarray}

\section{Band Structure}
We consider a dispersion, $\epsilon_{\textbf{k},b}$, that has two bands~(indexed by $b=\pm1$) centered at $b \epsilon_0/2$ with bandwidths $\max_{\textbf{k}}\left|\epsilon_{\left|\textbf{k}\right|}-b\epsilon_0/2\right|=W/2$, and a constant density of states within the two bands (see Fig.~\ref{fig:bandStructure}). Such a constant density of states can occur with Dirac cones and is required to be easily simulated by a uniform distribution for the inhomogeneous broadining on the atoms (see Fig.~\ref{fig:bandStructure}). {Furthermore, we work in the limit $W<2\epsilon_0$ such that there are two bands separated by a band gap of $\epsilon_0-W/2$. Note that, while the energy difference between the atomic excited state and its ground state is $2\epsilon_i$, each atom represents two fermion modes with momentum $-\textbf{k}$ and $\textbf{k}$ such that $\epsilon_i\rightarrow \epsilon_{\textbf{k},b}$ is the energy of a fermion at either of those momenta.
\begin{figure}[h!]
    \centering
    \includegraphics[height=1.26in,width=2.5in]{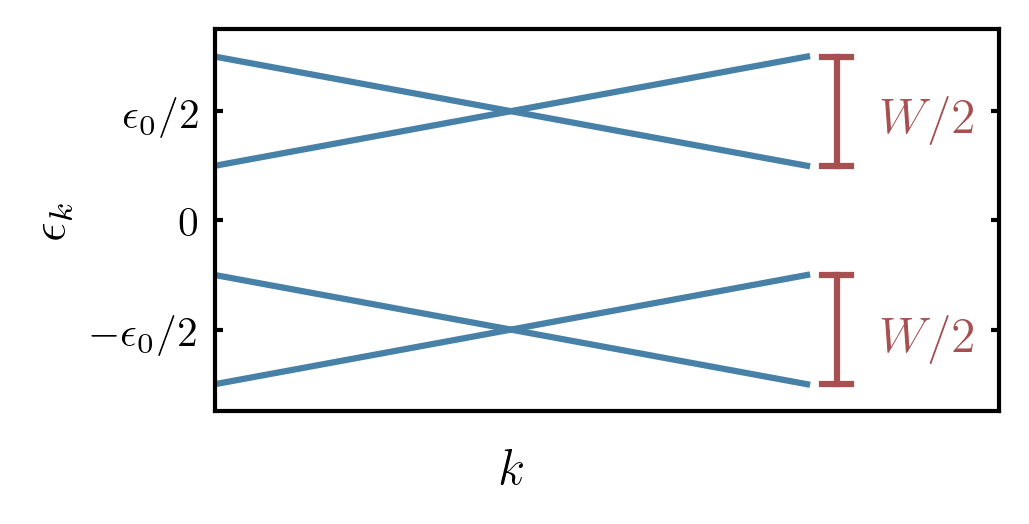}
    \caption{ An example band structure with density of states matching the distribution of the inhomogeneous atomic energies shown in the left panel. In this diagram, $\epsilon_0>W/2$ such that there is a band gap of size $\epsilon_0-W/2$. Note that center of the bands are at $\pm\epsilon_0/2$ for the fermion band structure, and not $\pm\epsilon_0$ as for the atomic level distribution. This is because there are two fermions per cooper pair, and it is cooper pairs that map to atomic levels.}
    \label{fig:bandStructure}
\end{figure}

\section{Equilibrium Phase Diagram}
While in the main text we consider quench dynamics from a state in which every possible Cooper pair is condensed, intuition about superconducting systems is often developed near thermal equilibrium.
To make connection to that limit, we consider the superconducting order of the grand canonical state $\rho_{\mu,T}=\exp[-({H-2\mu(S^z+a^{\dagger}a)})/{K_bT}]$ with chemical potential $\mu$ and temperature $T\rightarrow 0$, and
\begin{eqnarray}\label{eq:smham}
    H/\hbar =\sum_{i=1}^{2N} \epsilon_i \sigma^z_i +\sum_{i=1}^{2N} g_i (\sigma^{\dagger}_i a+ h.c.) + \Delta_c a^{\dagger}a.
\end{eqnarray}
This model has a $U(1)$ symmetry generated by the conservation of total excitations $\sum_i\frac{1}{2}\sigma^z_i+n$. 
At high $T$ and $\mu=0$, the system enters a symmetric phase with $\left<\vec{\sigma}\right>\rightarrow 0$ and with photon occupation scaling monotonically with $K_b T$.
At low temperature, a symmetric phase   with all spins polarized on the $\hat{z}$ axis, $\left<\sigma^z_i\right>=\text{sign}(\epsilon_i-\mu)$ (a normal state insulator or conductor), competes with a symmetry broken phase can where $\left|\left<S^+\right>\right|>0$ and $\left<a\right>>0$.
Away from the quantum critical point separating these two phases, a   mean field approximation (i.e. $\left<O_1(t)O_2(t)\right>=\left<O_1(t)\right>\left<O_2(t)\right>$) is valid, and we   again adopt the notation:~$\left<O_1\right>\equiv O_1$.
At zero temperature, energy is minimized at the fixed points of the classical dynamics, $\partial_t \vec{\sigma}=0$ and $\partial_t a=0$.
The later condition fixes the photon in a way mathematically equivalent to the adiabatic condition:
\begin{eqnarray}\label{eq:grnda}
    a= -\frac{g}{\Delta_c-2\mu}\sum_i \sigma^{-}_i(t).
\end{eqnarray}
Defining the gap as $\Delta=\frac{g^2}{\Delta_c-2\mu}\sum_i\sigma^+_{i}$ and inserting Eq.~\ref{eq:grnda} into Eq.~\ref{eq:smham}, one finds that the spins minimize energy in an effective magnetic field  $\vec{h}=(\epsilon_i-\mu)\hat{z}+\text{Re}(\Delta)\hat{x}+\text{Im}(\Delta)\hat{x}$.
Minimizing the energy of the spins in this field we find:
\begin{eqnarray}
    \sigma^{+}&=&\frac{\Delta}{2\sqrt{(\epsilon_i-\mu)^2+\Delta^2}} \\ \nonumber
    \sigma^{z}_i&=&\frac{\epsilon_i-\mu/2}{\sqrt{(\epsilon_i-\mu)^2+\Delta^2}}
\end{eqnarray}
using $g^2=\chi N\Delta_c$ we get the gap equation:
\begin{eqnarray*}
    1=\frac{\chi N \Delta^c}{W(\Delta^c-2\mu)}\int_{\mathcal{R}}d\epsilon\frac{\Delta}{\sqrt{(\epsilon-\mu)^2+\Delta^2}}
\end{eqnarray*}
where the integral is over the region defining the constant density of states: $\mathcal{R}=\epsilon\in (\epsilon_0/2-W/4,\epsilon_0/2-W/4) \cup (-\epsilon_0/2-W/4,-\epsilon_0/2-W/4)$.
In addition to the gap equation, we can also restrict the total number of excitations to match that of the initial state in the paper:
\begin{eqnarray}
    a^{\dagger}a+S^{z}=0=\frac{N}{2W}\int_{\mathcal{R}}d\epsilon\frac{\epsilon-\mu}{\sqrt{(\epsilon-\mu)^2+\Delta^2}}+\frac{\Delta^2}{\chi \Delta_c}
\end{eqnarray}
In the adiabatic limit, $\Delta_c/(\chi N)\gg1$, the photon amplitude goes to zero, and thus $S^z$ must also limit to zero.
This is ensured by a chemical potential $\mu=0$ such that $\epsilon_i-\mu$ is positive for half the spins and negative for the other half.
The resulting zero temperature phase diagram is calculated numerically and is shown in Fig.~\ref{fig:SMeq}.  At finite temperature, the superconducting coherence $\left|S^{+}\right|$ is reduced in the ordered phase and remains $0$ in the disordered phase.
Weather the disordered phase is a insulator or conductor at low temperature depends on if the chemical potential $\mu=0$ is located in a band or band gap.
Upon consideration of Fig.~\ref{fig:bandStructure}, we see the disordered phase is a two-band insulator when $\epsilon_0>W/2$ and a single band conductor for larger dispersion $W$.
Upon inspection of Fig.~\ref{fig:SMeq}, we see that for the adiabatic limit studied in the paper, the system would traditionally be considered a two-band insulator.

\begin{figure}[h!]
    \centering
    \includegraphics[height=2.5in,width=2.5in]{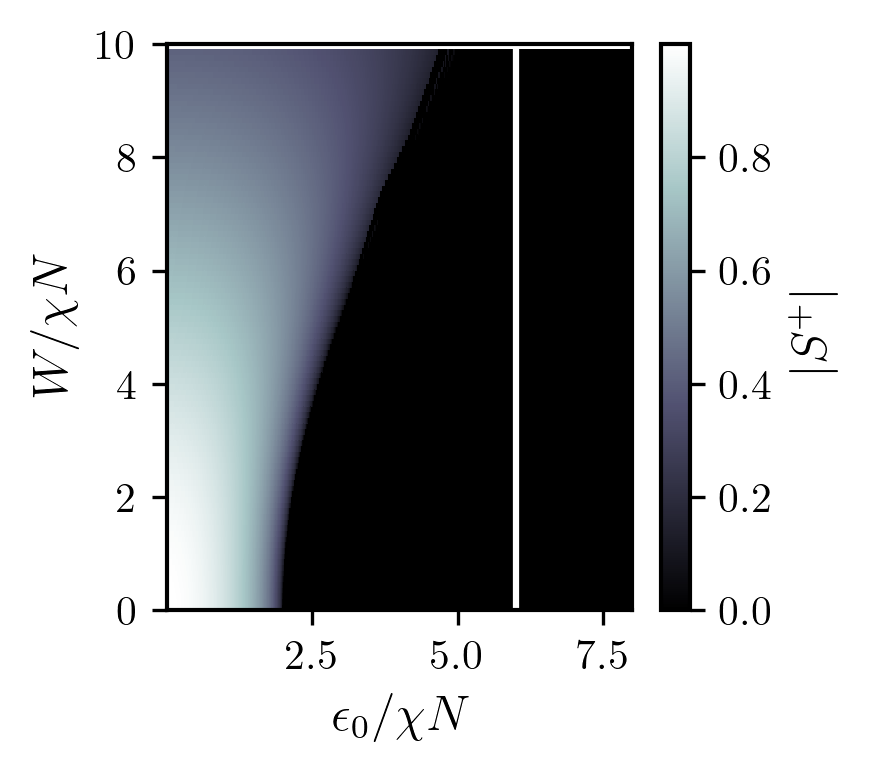}
    \includegraphics[height=2.5in,width=2.5in]{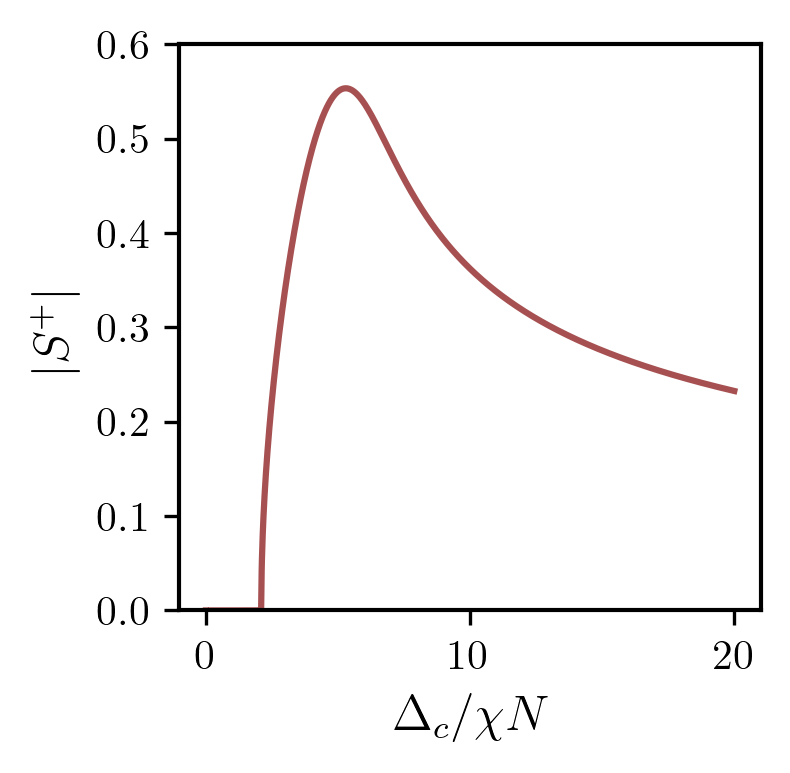}
    \caption{ \textit{Left Panel}: Equilibrium phase diagram at $T=0$ and with chemical potential fixed to $\mu=0$.  In the paper we study the line along $\epsilon_0/\chi N=6$ which has $\left|S^{+}\right|=0$ in the ground state and at all finite temperatures. \textit{Right Panel}: Zero temperature coherence, $S^{+}$, versus photon detuning, $\Delta_c/\chi N$, with $\mu$ chosen to fix the total number of excitations, $a^{\dagger}a+\frac{1}{2}\sum_i \sigma^z_i=0$. This panel is for, $\epsilon_0/\chi N=6$ and $W/\chi N=8$ as in the paper and shows a similar resonance peak.
    }
    \label{fig:SMeq}
\end{figure}

The phase diagram shown in the left panel of Fig.~\ref{fig:SMeq} holds for all values of $\Delta_{c}/\chi N$ assuming the chemical potential is fixed to $\mu=0$. If, instead, $\mu$ is chosen to fix the total number of excitations, it will depend on the detuning of the photon since, in the ordered phase, the photon occupation increases monotonically with $\chi N/\Delta_c$. Therefore, to maintain a fixed number of total excitations, the chemical potential must increase to ensure a finite and negative $S^{z}=-a^{\dagger}a$.  This can lead to instabilities and produce an ordered phase where there was none in the adiabatic limit.  Fixing $a^{\dagger}a+\frac{1}{2}\sum_i\sigma_i^z=0$, we find a similar resonant order at zero temperature as was seen out-of-equilibrium (see right panel of Fig.~\ref{fig:SMeq}).  Note, such an equilibrium state is unlikely to be observed in the cavity QED system due to the effects of dissipation which lead to a disordered state with no photons in the cavity at long times.

\section{Lindblad Dynamics}
In order to investigate the effect of photon loss, we study dynamics of the spin and cavity evolving under the Liouvillian with jump operator $L=\sqrt{\kappa/2}a$.
In the Heisenberg picture operators evolve following
\begin{eqnarray}\label{eq:smlindblad}
    \partial_tO=i\left[H,O\right]+2L^{\dagger}O L-\left\{L^{\dagger}L,O\right\} 
\end{eqnarray}
where $\kappa$ is the loss rate. In the mean field limit (i.e. $\left<O_1(t)O_2(t)\right>=\left<O_1(t)\right>\left<O_2(t)\right>$), we can truncate the hierarchy of equations generated by Eq.~\ref{eq:smlindblad}.
We then numerically evolve the closed set of equations for the spin and photon variables.
\end{document}